# Underprepared for Physics: Reframing the narrative on readiness and instruction in calculus-based, introductory physics courses

*Suzanne White Brahmia and Geraldine L. Cochran*

Isaac, the valedictorian of his urban high school, aspired to be an engineer. He excelled in every available math and science class, but his school didn't offer calculus. After enrolling in his state university, a math placement exam placed him into precalculus, making him ineligible for the calculus-based physics and chemistry courses required for engineering. Told this would delay graduation by at least a year, Isaac pushed forward--but felt increasingly disconnected from the engineering track. The added cost of a fifth year ultimately led him to switch majors, choosing a path he could complete in four years without overwhelming his family financially.

Isaac's story is not unique. To start taking physics courses in the U. S.--a common entry point for a mathematics-based career path like physics, computer science, and engineering--students typically must be enrolled in, or have completed, calculus. But this rigid requirement disproportionately excludes students from low-socioeconomic status (SES) districts where access to advanced math is limited. This exclusion is especially troubling given how little calculus is actually used in most introductory physics instruction.

Discussions around success in calculus-based physics often focus on *student readiness*--defined narrowly by prior calculus--while giving less attention to how well departments support the students admitted to their institution. Students labeled underprepared are typically required to complete remedial math, extending time to degree and increasing costs. Some persist; others are advised to change majors. Physics advising often reduces to a math test cutoff--pushing away capable students for reasons unrelated to their potential.

In this article, we examine evidence of unequal access to advanced mathematics before college, explore the unintended gatekeeping function of placement tests, and reflect on the skills that are actually necessary for success in introductory physics. As a compelling alternative to current practice, we highlight a long-running, successful program at Rutgers University that expands access while strengthening the integration of calculus concepts into physics. Looking forward, the article describes a national consortium that is beginning to coordinate resources and support departments working to improve outcomes for all students in introductory physics sequences.

## 1. Designating students as underprepared for calculus-based physics

Who gets to take physics in college often depends less on ability than on access to math opportunities long before college begins.[1] Pre-college education in the U.S. is marked by unequal access to advanced coursework, particularly in mathematics and physics. These gaps are shaped by broader structural inequities across school districts, often tied to poverty.

As shown in Figure 1, students attending high-poverty schools are significantly less likely to have an option to take calculus in high school. This opportunity gap is substantial, affecting the nearly *one-quarter* of U.S. public high school students who attend low SES schools, where at least 75% of students qualify for free or reduced-price lunch.[2,3]

**Figure 1:** U.S. High school calculus enrollment as a function of poverty concentration. Size of the dots reflects the school district size, and color represents the percentage of students who are underrepresented in STEM fields. (graph courtesy of M. Marder, 2021-2022 Civil Rights Data Collection[3])

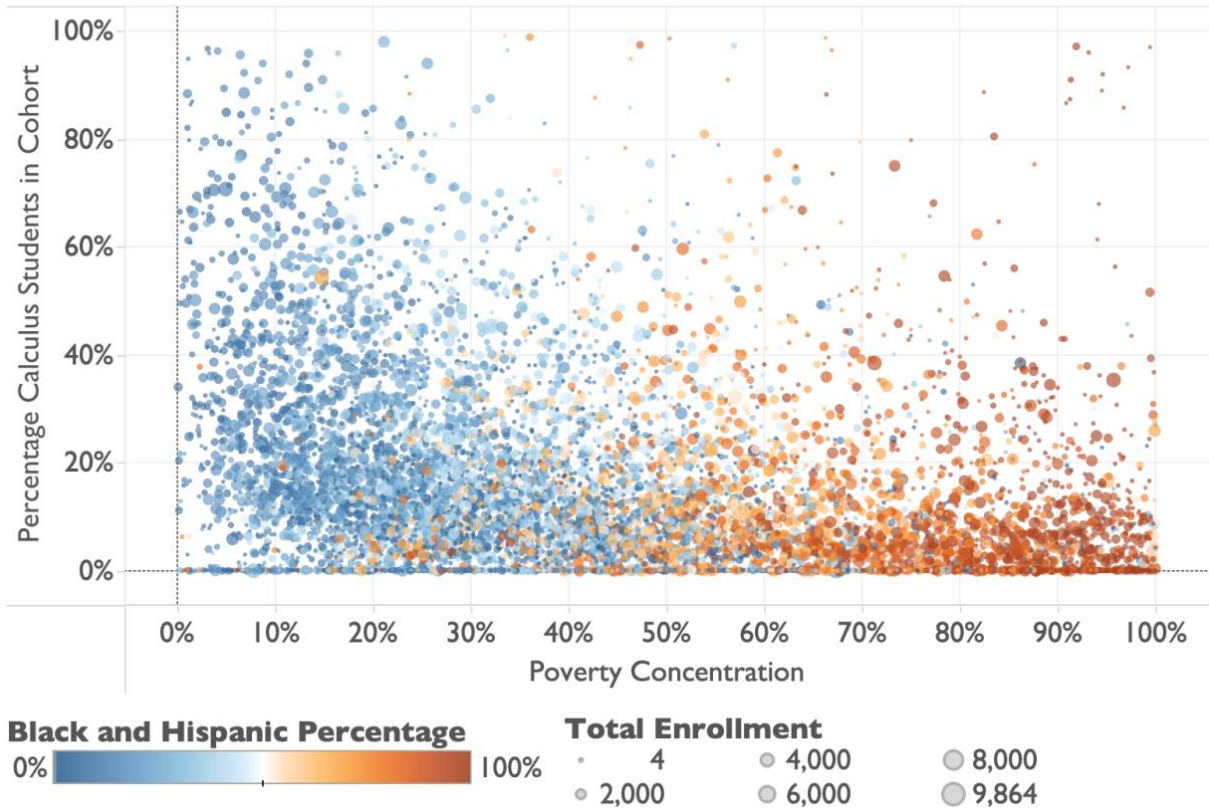

The disparities are even more pronounced for Black and Latino students.[3] In 2021, only 35% of public high schools with high Black and Latino enrollment offered calculus, compared to 54% of schools with lower enrollment of these groups.[4] That same year, only 86% of U.S. public high schools offered Algebra I at all.[4] As shown in Figure 2, Black students were nearly twice as likely as white students to attend a high school where calculus wasn't offered.[3]

These differences in course availability are not merely academic--they shape college trajectories and limit access to STEM majors, where calculus is often assumed as a prerequisite. Recognizing this context is essential for designing physics instruction and placement practices that do not penalize students for unequal access to opportunity.

Using math placement tests to determine readiness for physics courses, mirrors–and reinforces– inequities in educational opportunity. These assessments tend to promote the fundamental attribution error: interpreting a student's lack of calculus preparation as a personal shortcoming rather than a result of systemic barriers, such as unequal access to advanced math in high school.

**Figure 2:** Probability of no calculus-taking opportunities for US public school students identifying primarily as Black, Latinx, or White. (graph courtesy of M. Marder, 2020-2021 Civil Rights Data Collection[3])

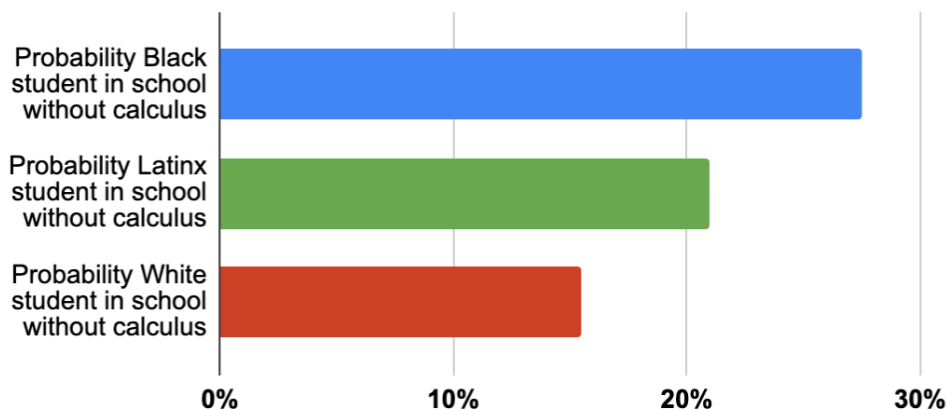

Compounding the issue, most placement tests emphasize procedural skills in algebra and trigonometry, for example:

Solve for x: (i) $4 + 5^x = 1000$ (ii) $\log_4(x - 3) = -2$

This kind of procedural competence is important, but drilling down on it uniquely as a barrier to entry is not aligned with the kinds of reasoning physics requires. Physics success relies more on *physics quantitative literacy* (PQL): the ability to interpret equations, apply math in context, and connect it to physical meaning[5,6]. This flexible, context-based reasoning is rarely taught in standard prerequisite math courses, yet it benefits all students--regardless of prior preparation[6]. While students who struggle with foundational algebra will need added support, equating high placement test scores with readiness for physics is deeply flawed.

Still, placement scores often serve as rigid gates, filtering out capable students and reinforcing opportunity gaps. This reflects a "broken-student" narrative--the idea that students must fix themselves to belong--when in fact, many were never given a fair opportunity to begin with.

Even among those who do enroll, disparities in preparation shape outcomes. About 75% of students who place into college calculus took it in high school, putting those without that opportunity at a disadvantage.[7] A study by Salehi et al.[8] across three selective institutions found that physics exam scores correlated with math SAT scores (Fig. 3) and prior physics experience--both tied to family income[3,9].

However, when course design aligns with student preparation, performance gaps shrink. Stewart et al.[10] found that controlling for SES or SAT scores largely eliminated ethnic disparities in *learning gains* – the actual learning during a course. Rather than asking who is prepared for physics, we should ask whether our courses are prepared for the students our institutions enroll.

**Figure 3:** SAT Math scores vs family income (from College Board 2009[9]).

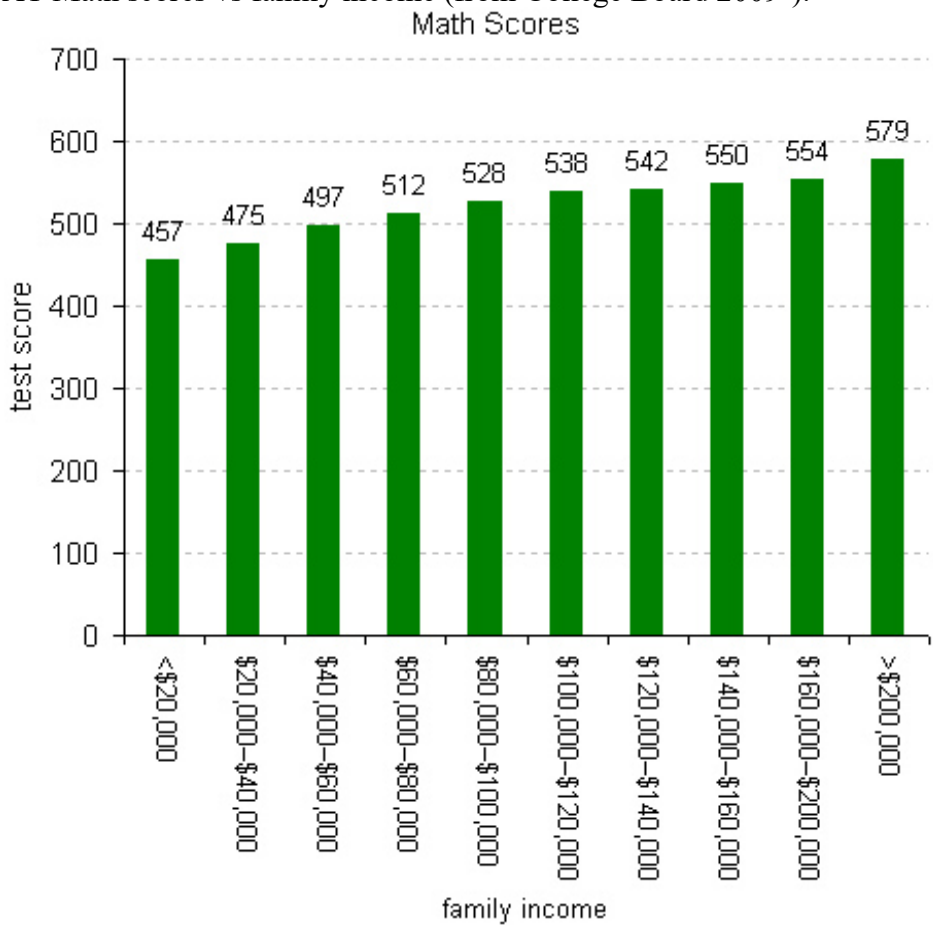

## 2. Preparing physics departments to teach all its students

Focusing only on student preparation–rather than the effectiveness of instruction at supporting diverse learners–can have unintended consequences. It often relies on metrics misaligned with the goals of physics courses, reflects disparities in prior access more than ability, and disproportionately affects students historically excluded from STEM. It can also lead to students questioning whether they belong in physics at all.

Remediation-heavy approaches often place the burden on students without addressing the underlying issues in course structure. Even valuable, well-intentioned supports like tutoring or bridge programs require extra effort from students, while leaving the courses themselves unchanged. A more effective approach focuses on redesigning instruction to support a broader and more diverse range of learners. Instead of requiring students to complete remedial math before enrolling in physics, departments can embed *physics quantitative literacy* (PQL)--the contextual, applied math skills that physics truly demands--directly into the physics sequence. This integration helps students develop the ability to interpret equations, reason with physical quantities, and connect mathematical relationships to real-world phenomena. Optional instructor-led support courses or extended, credit-bearing pathways that integrate PQL into instruction offer a more inclusive and effective alternative. They improve outcomes not only for students with less traditional preparation but also for those navigating the broader conceptual challenges of physics.

A sensible starting point for integrating PQL support is to examine how we use math in introductory physics. What does PQL involve that makes it different from a mathematics course? This question is a vibrant area of inquiry in Physics Education Research. Most U.S. students do little actual calculus problem-solving in introductory physics courses--even ones designated as calculus-based[11]. Yet, reasoning about core calculus ideas--like variation, rate of change, and accumulation--in the context of over 100 new physics quantities like force, momentum or energy is essential! Mathematics with physics quantities is substantially more difficult for all new learners than the context-free practice students gain in a math course[5,12]. These conceptual skills are rarely outcomes of traditional calculus instruction, which tends to focus on symbolic manipulation for solving mathematical problems[12]--most of which are irrelevant in physics. Moreover, the math structures we depend on--basic operations with simple function types like linear and inverse proportionalities and quadratic polynomials--are more widely accessible to our students than advanced techniques like integration by partial fractions. By emphasizing how we construct and symbolize physical quantities, and their relationships to each other, we can better support all students in developing meaningful mathematical reasoning in physics.

For physics instructors, the lesson is clear: by identifying when students need specific skills associated with PQL and weaving them into instruction, we can boost learning without lowering expectations[6]. Programs using this approach report improved outcomes for all students[13]. Many departments already offer enriched instruction for top scorers through honors programs--why not invest similarly in students who lacked access to pre-college physics or calculus? The challenge isn't fixing students--it's designing courses that help all of them thrive. The Extended Course model does this by removing the calculus prerequisite and adding credit hours to develop PQL in real time, alongside core physics content. Rutgers' Extended Analytical Physics program demonstrates how this can work in practice.

## 3. Case Study: Rutgers' Extended Analytical Physics

Rutgers, The State University of New Jersey, is a diverse, urban research institution. Since 1986, its Department of Physics and Astronomy has supported mathematically underprepared engineering students through the Extended Analytical Physics (EAP) program.[14,15,16] Launched with state and federal funding, the program aimed to address the disconnect between New Jersey's diverse population and the demographics of STEM graduates at its flagship university. Pre-college calculus enrollment in New Jersey public schools mirrors the national trend (Fig. 1), occurring primarily in affluent districts with few Black and Latino students. To address this disparity, Rutgers created a parallel physics pathway that allows students not placed into calculus to stay on track for engineering degrees.

**Figure 4:** Progression of the Extended Analytical and Analytical Physics pathways[14]

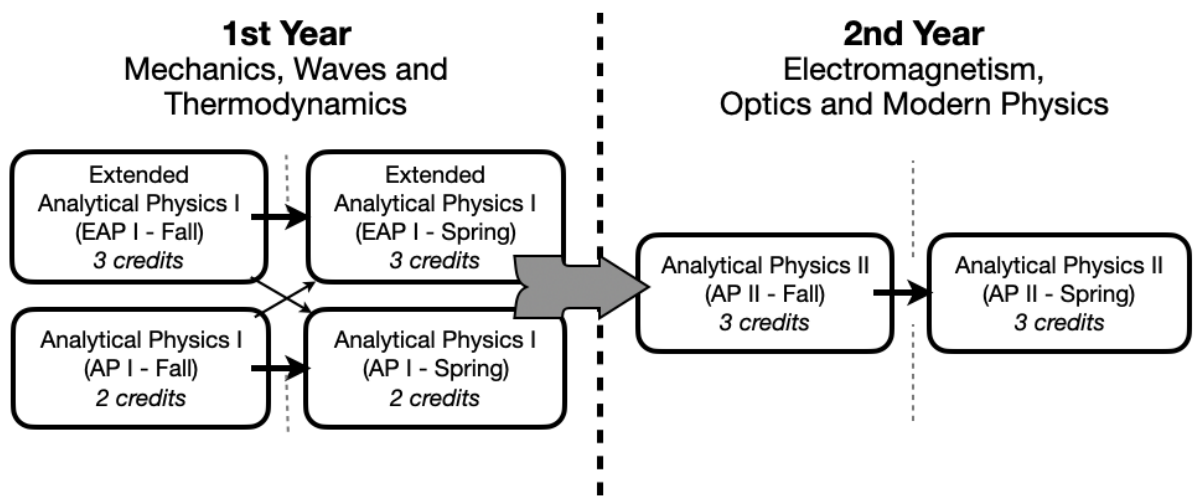

The EAP pathway (top pathway, Fig. 4) is an introductory physics sequence for engineering majors who place below calculus[14]. It spans three or four semesters, totaling 9 to 12 credit hours depending on the major. Typically, students take EAP I in fall and spring, followed by Analytical Physics II (AP II) in the fall--and for some majors, a second AP II course in the spring (top pathway, Fig. 4). This alternative complements the standard Analytical Physics (AP) sequence, which also runs three or four semesters with 7 to 10 credits (bottom pathway, Fig. 4). While the AP sequence reduces first-year physics credit load, EAP spreads the same content over more credits each semester, providing additional time and support. Since its launch, EAP enrollment has grown from 90 to approximately 300 students annually (as of 2021). Most students follow a consistent sequence, though some switch pathways. A separate honors track exists but is not shown.

EAP exemplifies an extended course structure, granting students an extra credit hour each semester for increased contact time and deeper engagement with physics concepts and physics quantitative literacy (PQL). Importantly, the course does not teach remedial math; instead, it helps students understand how algebra, precalculus, and introductory calculus concepts apply within physics contexts, introducing topics as needed. The program has broadened access to STEM degrees for students from diverse educational backgrounds.

Figure 5 illustrates EAP's impact by comparing two periods--before EAP's inception and seven years after--averaged over two years to smooth fluctuations[14]. The chart shows percentages of all students, female-identifying students, and students from minority ethnic/racial groups who passed first-year physics (regardless of pathway) and earned STEM degrees within six years. A conservative estimate of uncertainty is about 4%.

The program is meeting its objectives. Degree completion for all students is boosted by the remarkable gains among female students and those from historically marginalized groups in engineering. Notably, since EAP's implementation, the percentage of underrepresented minority (URM) students completing STEM degrees within six years has increased by over 40%. A ten-year follow-up study on first-year passing rates and subsequent AP II grades yielded similar results.

**Figure 5:** First year passing rate and degree completion for all engineering students, and for underrepresented groups, averaged over two years before, and seven years after the introduction of the Extended Analytical Physics option[14]

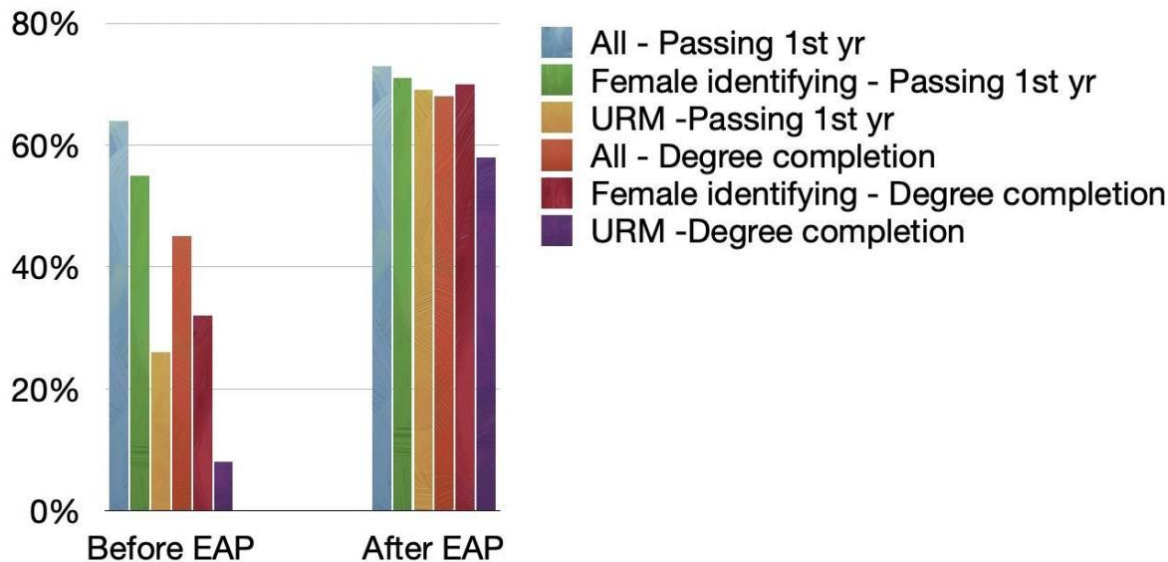

The strength and longevity of the EAP model lie in implicit structures that build student agency within a rigorous scientific community:

- **Flexible Entry:** Placement scores advise students, but they can choose or switch pathways through the start of spring semester, maintaining control over course placement.

- **Representative Instructors:** Lead faculty include members from groups underrepresented in physics, providing vital role models.

- **Supportive Environment:** The program fosters a safe pedagogical space where students can take risks and learn from mistakes.

- **Deep Learning Focus:** Activities emphasize conceptual and procedural understanding of linear and inverse proportional relationships, extending this reasoning to other key functions common in physics models.

- **Calculus Foundations, Without Calculus:** Students explore core calculus ideas-- quantities, rates of change, accumulation--through accessible precalculus reasoning.[12]

The Rutgers EAP model integrates PQL development into standard introductory physics by emphasizing quantitative reasoning rarely addressed in math courses but essential for physics.

Physical quantities--central to every physics model--relate through a few core equation types recurring across contexts. Helping students identify the mathematical role of each quantity--as change, rate, interval, or accumulation--deepens understanding of precalculus concepts and prepares students to engage with the scientific ideas those quantities represent. Crucially, these reasoning forms are accessible to precalculus students, focusing on conceptual rather than procedural calculus skills.

Developing PQL also means interpreting symbols and letters as representations of measurable, variable quantities--with units, and often with direction or sign. Vector quantities add representational complexity requiring fluency with notation like unit vectors, subscripts, and signed scalars. These conventions convey essential information about orientation and reference frames, vital for accurately modeling physical systems and suitable for introduction before formal calculus.

The Rutgers EAP program serves as a model for effective expanded access and sustained success. While some institutions of higher education are beginning to rethink introductory physics through an access lens, progress remains limited. Given the disproportionate impact of the pandemic on students from low SES backgrounds, it is critical to reevaluate access criteria for calculus-based physics and expand programs effectively supporting these students. The following section highlights a national initiative addressing this challenge.

## 4. Supporting efforts to support all capable students

The TIPSSS network (Transforming Introductory Physics Sequences to Support all Students)[17] is in its early stages of development, helping connect departments and educators committed to rethinking introductory physics instruction for all driven, capable students, regardless of what math course they had the privilege of taking in high school.

Funded by the NSF, TIPSSS supports departmental transformation, adapts modular curricula, and studies student learning and identity[18]. Its resources promote PQL, helping instructors customize materials for their students. TIPSSS also offers a rare professional community for instructors driving change. TIPSSS is a step toward collective action--connecting departments committed to rethinking instruction and broadening access so physics becomes a path, not a barrier, to students' futures.

Meeting students where they are requires rethinking long-standing course designs with sustained effort and institutional support. Research on PQL and programs like Rutgers' EAP show that improvement is possible. Physicists are natural problem solvers, but we cannot single-handedly fix the deep disparities in U.S. pre-college math education. That essential work is underway elsewhere and will take time. Meanwhile, we have agency. As post-secondary faculty, we can rethink the signals we send through course design and placement policies. Physics instructors share a commitment to unlocking student potential. Now, we must ensure our instruction supports all students--not just those fortunate enough to take physics and calculus in high school. Isaac's story may be common, but it doesn't have to be the norm. What are we doing to make sure students like Isaac aren't turned away before they've had a real chance to pursue the futures they envision?

# Bibliography

1. Schmidt, W. H., Burroughs, N. A., Zoido, P., & Houang, R. T. (2015). Educational researcher, 44(7), 371-386.

2. NCES. (2024). U.S. DoED IES. https://nces.ed.gov/programs/coe/indicator/clb.

3. Marder, M., visualizations using data from 2020-2021 and 2021-2022 Civil Rights Data Collection
https://institute.uteach.utexas.edu/profiles/michael-marder

4. U.S. DoED, Office for Civil Rights,(2023). https://civilrightsdata.ed.gov.

5. Edward F. Redish, *Phys. Teach.* 1 May 2021; 59 (5): 314–318.

6. White Brahmia, S., Olsho, A., Smith, T.I., Boudreaux, A. Eaton, P. and Zimmerman, C., (2021), Phys. Rev. Phys. Educ. Res. 17, 020129

7. Bressoud, D., (2021), Univ. Texas Dana Center
https://www.utdanacenter.org/blog/decades-later-problematic-role-calculus-gatekeeper-opportunity-persists

8. Salehi, S. Burkholder, E., Lepage, G. P. Pollock, S. & Wieman, C., (2019), Phys. Rev. Phys. Educ. Res. 15, 020114.

9. College Board 2009 https://secure-media.collegeboard.org/digitalServices/pdf/research/cbs-2009-national-TOTAL-GROUP.pdf

10. Stewart, J., Cochran, G., , Henderson, R., Zabriskie, C., DeVore, S., Miller, P., Stewart, G., & Michaluk, L. , Phys. Rev. Phys. Educ. Res. 17, 010107.

11. Loverude, M.: Where is the calculus in calculus-based introductory mechanics – a textbook analysis. In: Proceedings of the 27th Annual Conference on Research in Undergraduate Mathematics Education (2025)
https://matriccalcconf3.sciencesconf.org/data/pages/Pre_proceedings_2.pdf

12. White Brahmia, S., Thompson, P., (2025). Int. J. Res. Undergrad. Math. Educ. *(in press)*.
https://doi.org/10.48550/arXiv.2501.04219

13. A M Capece and A J Richards 2024 Phys. Educ. 59 055014

14. White Brahmia, S. (2008). AIP Conf. Proc. 1064, 7–10.
https://doi.org/10.1063/1.3021279

15. Brahmia, S.& Etkina, E., (2001):, J. Coll. Sci. Teach.; Abingdon Vol. 31, Iss. 3, 183-187.

16. Holton, B. E., & Horton, G. K. (1996). *The Physics Teacher*, *34*(3), 138-34.

17. TIPSSS (*Transforming Introductory Physics Sequences to Support all Students)* Network. url: http//xxx. osu.edu

18. Research on identity formation in physics and engineering:
    - Patrick, A. D., & Prybutok, A. N. (2018). Int. J. Eng. Educ, 34(2a).
    - Hazari Z, Chari D, Potvin G, Brewe E. *J Res Sci Teach*. 2020; 57: 1583–1607.